\documentclass[11pt,floatfix]{JHEP3}
\usepackage{amssymb}
\usepackage{amsmath}
\usepackage{epsfig}

\preprint{}
\title{Dynamical decompactification from brane gases
       in eleven-dimensional supergravity}
\author{
Antonio Campos\\
Institut f\"ur Theoretische Physik\\
Universit\"at Heidelberg\\
Philosophenweg 16\\
69120 Heidelberg\\
Germany\\
E-mail: \email{a.campos@thphys.uni-heidelberg.de}
}
\abstract{
Brane gas cosmology provides a dynamical decompactification
mechanism that could account for the number of spacetime 
dimensions we observe today. 
In this work we discuss this scenario taking into account
the full bosonic sector of eleven-dimensional supergravity.
We find new cosmological solutions that can dynamically
explain the existence of three large spatial dimensions 
characterised by an universal asymptotic scaling behaviour
and a large number of initially unwrapped dimensions. 
This type of solutions enlarge the possible initial
conditions of the Universe in the Hagedorn phase and
consequently can potentially increase the probability of
dynamical decompactification from anisotropically wrapped 
backgrounds.
}


\begin{document}

\section{Introduction}
Understanding why we live in 3+1 dimensions is an intriguing
and puzzling problem in theoretical physics.
In the standard theory of general relativity the dimensionality
of the Universe is an assumption and it cannot be derived 
dynamically or from a fundamental law. 
A suitable theoretical framework to deal with this question
is provided by string theory which predicts the existence of
extra dimensions. 
The general lore is to assume that these dimensions are 
effectively very small and consequently unobservable today.
It is, however, fairly plausible to belive that this simple
picture changes in the far past.
As the Universe shrinks and gets smaller all directions 
should behave in the same footing and scale with similar 
sizes.
Until definite physical evidence of the contrary is put
forward one can consider the very early Universe to be higher 
dimensional and try to unveil a dynamical process responsible
for the late asymmetry between large and small spatial 
directions.

An important ingredient of string theory is the presence
of extended objects.
The idea that these objects can play a fundamental role in 
explaining the current number of dimensions was firstly suggested 
by Brandenberger and Vafa \cite{Brandenberger:1989aj} (see also 
\cite{Tseytlin:1992xk,Tseytlin:1992ss,Kripfganz:1988rh,Bassett:2003ck}).
In this proposal the background spacetime is considered
to have (d+1)-dimensions with a spatial toroidal section
and the dynamics is assumed to be driven by a gas of fundamental
strings. 
Basically, all spatial directions are initially small, namely 
with a typical size of the order of the string length.
As the Universe expands the energy of the winding modes that
appear in the spectrum increases leading to a confining potential
that stops the expansion. 
Then, to have a large Universe one has to devise a way to
get rid of these modes.
An efficient annihilation mechanism of string winding modes can 
only take place if the dimensionality of the spacetime is smaller 
than 4.
This can be very easily understood on topological grounds.
If the dimensionality of the spacetime is higher the probability of 
interaction of one dimensional extended objects is basically 
negligible.
The picture is that of a Universe starting with all directions
oscillating independently around the string length scale until 
incidentally three of them get slightly larger than the others. 
Then, the process of annihilation is triggered letting the 
Universe grow with the appropriate dimensions. 

Recently, it has been proved that this mechanism also works 
if the dynamics of D-branes is taking into account 
\cite{Alexander:2000xv}.
The key argument is again topological because the probability 
of interaction of D-branes depends basically on the 
dimensionality of their worldvolumes.
The larger the dimensionality the most effective the decay
process is.
In this picture, the dimensionality of the background spacetime 
is successively  lowered as branes with larger dimensions are 
annihilated.
Since the last winding modes to decay are those of the 
one-dimensional D-branes, an expanding Universe with three large 
dimensions and a hierarchy of small dimensions can be explained.
Some particular aspects of the cosmology of brane gases have
been studied by different authors
\cite{
Brandenberger:2001kj,Campos:2003gj,Watson:2002nx,
Boehm:2002bm,Alexander:2002gj,Kaya:2003py,Kaya:2003vj,
Brandenberger:2003ge,Campos:2003ip,Biswas:2003cx,
Watson:2003uw,Watson:2004vs}.
An interesting question that arises is whether brane
gases can naturally stabilise the small extra dimensions 
at the string scale.
In fact, this has been tested and verified in different
set ups
\cite{
Watson:2003gf,Battefeld:2004xw,Kim:2004ca,Watson:2004aq,
Kaya:2004yj,Patil:2004zp}.
However, as it has been pointed out in \cite{Berndsen:2004tj},
it seems quite hard to stabilise simultaneously the volume 
of the internal space and the dilaton without invoking new 
physics.

The problem of dilaton stabilisation also brings a conceptual 
drawback.
The existence of the dilaton itself is a consequence of 
assuming {\sl a priori} that one of the dimensions is already 
compactified on a circle.
Then, to have a proper mechanism of dynamical decompactification
one should have started from a more fundamental theory.
This was the purpose of the framework proposed in 
\cite{Easther:2002qk} for studying the cosmology of brane 
gases.
Here, the starting point is the purely gravitational sector
of eleven-dimensional supplemented with a gas of massless 
supergravity particles and a gas of nonrelativistic 
M2-branes wrapping anisotropically the spatial directions.
Dimensions which are unwrapped become large at late times
because of the absence of negative pressures that could 
suppress their expansion.
Then, this scenario can explain a proper hierarchy of 
dimensions if there are initially three unwrapped spatial 
directions.
Although a hierarchy of dimensions appears rather naturally 
the small dimensions do not stabilise in the relevant cases.
As noted in \cite{Alexander:2002gj}, a decompactification 
mechanism with isotropic wrapping can be also obtained if the 
intersections between M2- and M5-branes, which represent 
string-like degrees of freedom, are included in the dynamics.

How plausible anisotropically wrapped configurations are depends 
on the thermodynamics of the brane gas close to the Hagedorn 
temperature.
The analysis of this phase reveals that anisotropic wrappings with
a low number of unwrapped dimensions are only compatible with a 
large initial volume of the Universe \cite{Easther:2003dd}.
Basically, this is a consequence of the fact that for small volumes
the rate of annihilation of branes increases and a larger number
of dimensions can get unwrapped more rapidly.
However, to put bounds on the initial conditions of the Universe 
and see whether a number of unwrapped dimensions is preferred at 
the time in which branes and antibranes freeze out new arguments
are needed.  
A speculative possibility is to impose that the very early Universe 
is consistent with the holographic principle.
This principle requires that the entropy inside a spherical volume 
must be bounded by the surface area \cite{Bousso:1999xy,Bousso:1999cb}. 
Estimating the entropy density close to the Hagedorn phase one 
can check that the initial state of the Universe should be rather
small and, then, initial brane configurations with a small
number of unwrapped directions are strongly disfavoured
\cite{Easther:2003dd}.
Nevertheless, this conclusion was obtained assuming isotropy 
of the background spacetime and should be taking very cautiously.

The main purpose of this work is to investigated how the 
dynamics of a gas of branes changes if the gauge sector of 
eleven-dimensional supergravity is also taking into account
and see whether some of the previous problems can be solved.
There are not many works analysing the importance of gauge
fields in the context of brane gas cosmology.
In \cite{Alexander:2002gj}, it has been noticed that fluxes 
can be responsible for the existence of a subhierarchy of 
small dimensions in scenarios with M2-M5 intersections, and 
in \cite{Campos:2003ip}, it has been proved that a gauge field 
can dominate over the confining potential of the winding modes 
at late times and be responsible for the expansion of three 
large directions in a ten-dimensional supergravity background.

Naively, one would expect that fluxes with a non-negligible
dynamical contribution to the dynamics at late times will
spoil the decompactification mechanism with anisotropic 
wrappings.
Generically, a cosmology driven by a 4-form gauge field strength
in a spatially flat background has seven expanding and three 
contracting spatial dimensions \cite{Demaret:1985js}.
Although a hierarchy of dimensions is dynamically created,
it does not predict the right number of large dimensions.
We will argue that in fact the dynamics of fluxes can be easily 
accommodated in this decompactification mechanism.
Furthermore, it is shown that {\sl the presence of fluxes 
allows a new class of cosmological solutions with a large 
number of unwrapped dimensions which can account for three 
large spatial dimensions at late times.}
Solutions with this type of brane configurations make less 
stringent the initial conditions of the Universe in the
Hagedorn phase and consequently enlarge the probability of 
dynamical decompactification from anisotropically wrapped 
brane gases.

\section{Brane gas dynamics in eleven dimensions}
We consider the eleven-dimensional $N=1$ supergravity
with the fermionic sector frozen out.
The degrees of freedom of this theory consist of a graviton 
and a 3-form gauge field $A_{[3]}$. 
The effective action is given by \cite{Cremmer:1978km}, 
\begin{equation}
S  = \frac{1}{2\kappa^2_{11}}
     \left[  \int d^{11}x\ \sqrt{-g}
             \left( R - \frac{1}{48} F^2_{[4]}
             \right)
           + \frac{1}{6}
             \int A_{[3]}\wedge F_{[4]} \wedge F_{[4]}
     \right]\, ,
\end{equation}
where $R$ is the scalar curvature, $F_{[4]}=dA_{[3]}$ is the field 
strength of the gauge field, and the eleven-dimensional gravitational 
coupling constant $\kappa_{11}$ is given in terms of the Planck length 
by $2\kappa^2_{11}=(2\pi)^8 l^9_{11}$.
In what follows, we will use Planck units in which $l_{11}=1$.
The last term is the Chern-Simons contribution and arises
as a direct consequence of supersymmetry.
In this work we only consider gauge fields with vanishing 
Chern-Simons term. 
For this class of solutions the classical equations of motion
are simply,
\begin{eqnarray}
G^{\mu\nu}
   & = &  \kappa^2_{11}
          \left( T^{\mu\nu}_{_G} + T^{\mu\nu}_{_M}
          \right) \, ,                                      \label{eq:Einstein}
\\
\nabla_\mu F^{\mu\nu\alpha\beta}
   & = & 0\, ,                                               \label{eq:Maxwell}
\end{eqnarray}
where $T^{\mu\nu}_{_G}$ is the energy-momentum tensor associated with
the gauge field,
\begin{equation}
T^{\mu\nu}_{_G}
   = \frac{1}{12\kappa^2_{11}} 
     \left( {F^\mu}_{\alpha\beta\gamma}
            F^{\nu\alpha\beta\gamma}
           -\frac{1}{8}g^{\mu\nu}
            F_{\alpha\beta\gamma\rho}
            F^{\alpha\beta\gamma\rho}
     \right)\, ,
\end{equation}
and $T^{\mu\nu}_{_M}$ stands for the energy-momentum tensor
of any other matter component.
Note also that because the field strength is an exact form 
the dynamics must also obey the Bianchi identity 
$\nabla_{[\rho} F_{\mu\nu\alpha\beta]}=0$.
Since we are interested to investigate under which conditions
three spatial dimensions grow large it is natural to consider
spacetimes which are homogeneous but spatially anisotropic,
\begin{equation}
ds^2
   = -dt^2 + \sum^{10}_{i=1} e^{2\lambda_i(t)} dx^2_i\, .
\end{equation}
We further assume that the spatial dimensions have the topology 
of a torus so that the spatial coordinates have a finite range 
$0 \leq x_i \leq 1$ and the total spatial volume is,
\begin{equation}
V 
   = \prod^{10}_{i=1}e^{\lambda_i}\, .
\end{equation}
The non-vanishing components of the Einstein tensor for this
metric are,
\begin{eqnarray}
G^t{}_t 
   & = & -\frac{1}{2} 
          \sum^{10}_{k \neq l} 
          \dot{\lambda}_k\dot{\lambda}_l\, ,
\\
G^i{}_j 
   & = & -\delta^i{}_j
          \left[  \sum^{10}_{k \neq i} 
                  \left( \ddot{\lambda}_k
                        +\dot{\lambda}^2_k
                  \right)
                - \dot{\lambda}_i
                  \sum^{10}_{k \neq i} 
                  \dot{\lambda}_k
                + \frac{1}{2}
                  \sum^{10}_{k \neq l}
                  \dot{\lambda}_k\dot{\lambda}_l
          \right]\, .
\end{eqnarray}
The equation of motion for the field strength~(\ref{eq:Maxwell}) 
can be straightforwardly solved using a Freund-Rubin ansatz
\cite{Freund:1980xh} (some cosmological solutions
have been investigated in 
\cite{Freund:1982pg,Alvarez:1984gi,
Demaret:1985js,Moorhouse:1985iz}
).
In this type of solutions the antisymmetric field strength is
considered to be nonzero only on a 3+1-dimensional submanifold,
say,
\begin{equation}
F^{\mu\nu\alpha\beta}
   = \frac{\epsilon^{\mu\nu\alpha\beta}}{\sqrt{-g_4}} F(t)\, ,
\end{equation}
where indices run from $0$ to $3$, $g_4$ is the determinant of the 
induced metric on the submanifold, $\epsilon^{\mu\nu\alpha\beta}$
is the ordinary Levi-Civit\`a density, and the function $F(t)$ is 
given by 
\begin{equation}
F(t)
   = f\, 
     e^{-\lambda_4(t)}\cdots e^{-\lambda_{10}(t)}\, ,     \label{eq:elementary}
\end{equation}
with $f$ a constant of integration.
The corresponding energy-momentum tensor for the gauge field is 
diagonal and its individual components can be compactly expressed as,
\begin{equation}
(T_{_G})^t{}_t 
   =  -\varepsilon^i (T_{_G})^i{}_i
   =  - \frac{1}{\kappa^2_{11}}
           \left( \frac{F(t)}{2}
           \right)^2\, ,
\end{equation}
where the ten-dimensional object $\varepsilon^i$ is
$-1$ for $i=1, 2, 3$ and $+1$ for $i=4, \cdots, 10$.
This energy-momentum tensor corresponds to a fluid with 
energy density, 
\begin{equation}
\rho_{_G} 
   = \frac{1}{\kappa^2_{11}}
           \left( \frac{F(t)}{2}
           \right)^2\, ,                              \label{eq:rho_elementary}
\end{equation}
and anisotropic pressures,
\begin{equation}
p^i_{_G} 
   = \varepsilon^i \rho_{_G}\, . 
\end{equation}
Note that as long as the gauge field has a dominant contribution
the spacetime is naturally separated into R$\times$T$^3\times$T$^7$.
The presence of the gauge field also introduces a new length scale 
into the dynamics given by the inverse of the initial vacuum 
expectation value of the antisymmetric field strength, 
\begin{equation}
l_{_G}
   \sim \left| \langle F_{\mu\nu\alpha\beta}
                       F^{\mu\nu\alpha\beta}
               \rangle
        \right|^{-1/2}
   \sim l^7_o f^{-1}\, ,
\end{equation}
where $l_o$ gives the typical length scale for the initial
size of the Universe. 
The gauge field will be dynamically relevant if $l_{_G}$ scales
with $l_o$ or, equivalently, if the integration constant of the 
gauge field strength is of the order $f\sim l^6_o$.  

Apart from the gauge field we still need to specify the rest 
of matter sources before trying to solve Einstein 
equations~(\ref{eq:Einstein}).
We assume that an important component of supersymmetric
matter is present in the early Universe.
This source of relativistic matter can be represented by a gas of 
massless particles, with energy density $\rho_{_S}$ and pressure 
$p_{_S}$. 
For simplicity we take the gas to be a homogeneous and isotropic
perfect fluid with a radiation equation of state 
$p_{_S} = \rho_{_S}/10$.
The corresponding energy-momentum tensor is,
\begin{equation}
(T_{_S})^\mu{}_\nu 
   = {\rm diag}(-\rho_{_S},p_{_S},\ldots,p_{_S})\, .
\end{equation}
Note that because this energy-momentum tensor is covariantly 
conserved, the energy density of supergravity particles 
scales with the total size of the Universe as,
\begin{equation}
\rho_{_S}
   = \rho^o_{_S} \left( \frac{V_o}{V}
                 \right)^{11/10}\, ,
\end{equation}
where $\rho^o_{_S}$ and $V_o$ are, respectively, the energy density 
and the spatial volume at some given time, $t_o$.

The second source of matter in this model is a gas of M2-branes 
wrapped on the various cycles of the torus.  
This gas can be characterised by a matrix of wrapping numbers 
$N_{ij}$, where elements with $i<j$ represent the number of branes 
wrapped on the $(ij)$ cycle while elements with $i>j$ represent the 
number of antibranes in the same cycle.
The elements of the diagonal are irrelevant and can be chosen
equal to zero.
Since we assume that the number of branes and antibranes is the
same for each cycle the wrapping matrix is symmetric.
For the later discussion it is useful to classify the spatial
dimensions into three types.
A direction $i$ is said to be {\sl unwrapped} if $N_{ij}=0$ for
all $j$.
{\sl Fully wrapped} directions have all $N_{ij}$ nonzero except
those that correspond to an unwrapped direction.
A direction for which some of the components $N_{ij}$ are
zero for values of $j$ corresponding to a not unwrapped direction
is referred to as {\sl partially wrapped}.
The actual values of this wrapping numbers should be provided 
by a, still lacking, theory of thermal and quantum fluctuations
in the very early Universe.
For our purpose, and until our theoretical understanding of this 
underlying theory is improved, we will simply take these numbers
as random integers.
Note that although the theory also supports the existence
of M5-branes, their dynamics is not taking into account 
because they decay very rapidly in eleven dimensions and
cannot produce any relevant effect at late times.

A single M2-brane is describe by the Nambu-Goto action,
\begin{equation}
S_{M2}
   = - T_2 \int_{{\cal M}_3} d^3\sigma\ \sqrt{-g_3}\, ,
\end{equation}
where the coordinates $\sigma^a (a=0,1,2)$ parametrise the
three-dimensional worldsheet manifold ${\cal M}_3$ spanned by
the brane and $g_3$ is the determinant of the pull-back onto 
${\cal M}_3$ of the eleven-dimensional bulk metric $g_{\mu\nu}$.
The surface tension of the brane is a fixed parameter given in 
Planck units by \cite{deAlwis:1996ez},
\begin{equation}
T_2
   = \frac{1}{(2\pi)^2}\, .
\end{equation}
Following \cite{Easther:2002qk} we will ignore excitations on the 
brane worldvolumes and assume that the branes are non-relativistic.
Under these conditions, the non-vanishing components of the stress 
tensor uniformly averaged over transverse directions for a brane gas 
with wrapping matrix $N_{ij}$ will be given by,
\begin{eqnarray}
\rho_{_B}=-(T_{_B})^t{}_t 
   & = &  \frac{1}{(2\pi)^2 l_{11} V} 
          \sum_{k \neq l} e^{\lambda_k} e^{\lambda_l} N_{kl}\, , 
\\
p^i_{_B}=(T_{_B})^i{}_i 
   & = & - \frac{1}{(2\pi)^2 l_{11} V} 
           \sum_{k \neq i} e^{\lambda_i} e^{\lambda_k}
           \left( N_{ki} + N_{ik}
           \right)\, .
\end{eqnarray}
where we have introduced the total energy density, $\rho_{_B}$, 
and the anisotropic pressures, $p^i_{_B}$, respectively.
The above M2-brane action assumes that the brane is moving in
the eleven-dimensional background without interacting with
the gauge field.
Nevertheless, it is easy to check that under our assumptions
coupling terms of the form \cite{Bergshoeff:1987cm},
\begin{equation}
 T_2 \int_{{\cal M}_3}A_{[3]}\, ,
\end{equation}
where $A_{[3]}(x(\sigma))$ represents the pull-back
of the eleven-dimensional gauge field 3-form onto ${\cal M}_3$,
do not contribute to the dynamics and can be safely neglected.

Now, we have to insert all the matter sources of energy-momentum, 
$T^{\mu\nu}_{_G}, T^{\mu\nu}_{_S}$, and $T^{\mu\nu}_{_B}$, into 
the right hand side of Einstein equations~(\ref{eq:Einstein}).
After some algebra the time component of the equations can be 
expressed as,
\begin{equation}
\sum^{10}_{k\neq l} \dot\lambda_k\dot\lambda_l
   =
    2\kappa^2_{11}
     \left( \rho_{_S} + \rho_{_B} + \rho_{_G}
     \right)\, ,                                          \label{eq:Einstein_t}
\end{equation}
and the spatial components as,
\begin{equation}
\ddot\lambda_i + 10 H \dot\lambda_i
   = \kappa^2_{11}
     \left[ \frac{1}{10}\rho_{_S}
           +\frac{1}{3}\rho_{_B}
           +\left( \varepsilon^i
                  - \frac{1}{3} 
            \right) \rho_{_G}
           +p^i_{_B}
     \right]\,\,\,\,\,\, i=1, \cdots, 10\, ,              \label{eq:Einstein_s}
\end{equation}
Here, $H$ is the mean Hubble parameter which represents the 
rate change of the total spatial volume of the Universe and is
given in terms of the metric components by,
\begin{equation}
H
   = \frac{1}{10}\sum^{10}_{i=1}\dot\lambda_i\, .
\end{equation}
One can readily observe that now there are two possible source 
of anisotropy in the cosmological evolution.
The one coming from the brane gas $p^i_{_B}$ and the one from 
the gauge field $\varepsilon^i \rho_{_G}$. 
A quantitative way of measuring the anisotropy of a spacetime 
is by means of the shear scalar which can be defined for our 
metric ansatz as,
\begin{equation}
\sigma^2(t)
   \equiv \frac{1}{9}
          \sum^{10}_{i=1} \left( \dot\lambda_i - H
                          \right)^2\, .
\end{equation}
Obviously, this object is zero only if all the expansion 
rates are equal. 
That the shear scalar can be nonzero in our model can be explicitly 
seen by comparing the evolution of two different expansion rates, 
\begin{equation}
 \ddot\lambda_i - \ddot\lambda_j
+10 H (\dot\lambda_i - \dot\lambda_j)
   = \kappa^2_{11}\left[ (\varepsilon^i-\varepsilon^j)\rho_{_G}
                        +(p^i_{_B}-p^j_{_B})
                  \right]\, .
\end{equation}
{}From this expression one easily observes that both the gauge field 
and the gas of M2-branes are the sources of a nonisotropic 
evolution of the Universe.
The pressures, $p^i_{_B}$, exerted by an anisotropically wrapped 
brane gas are always nonpositive and, then, their effect is
to suppress the growth of the scale factor in the dimensions
with nonzero wrapping numbers.
An asymmetric wrapping, in the sense that some spatial dimensions 
are unwrapped and then differential pressures are allowed, will 
lead to an anisotropic expansion.
On the other hand, the effect of the gauge field pressures,
$p^i_{_G}=\varepsilon^i\rho_{_G}$, depends on the sign of 
$\varepsilon^i$.
For those with $\varepsilon^i=+1$, the size of the spatial dimension
$i$ is enhanced and for those with $\varepsilon^i=-1$ suppressed.
Two spatial dimensions $i$ and $j$ will have different expansion 
rates if $\varepsilon^i\neq\varepsilon^j$.
Hence, it is of fundamental importance for understanding the asymptotic
cosmological evolution to see which energy components dominate at 
late times for different initial brane configurations. 
This is better done by introducing a new set of dimensionless functions 
of cosmic time,
\begin{eqnarray}
&&\Omega_\sigma(t) =  \frac{1}{10}\frac{\sigma^2}{H^2}\, ,
\>\>\>\>\>\>\>\>\>\>
\Omega_{_S}(t) =  \frac{\kappa^2_{11}}{45}\frac{\rho_{_S}}{H^2}\, ,
\\
&&\Omega_{_B}(t) =  \frac{\kappa^2_{11}}{45}\frac{\rho_{_B}}{H^2}\, ,
\>\>\>\>\>\>\>\>
\Omega_{_G}(t) =  \frac{\kappa^2_{11}}{45}\frac{\rho_{_G}}{H^2}\, .
\end{eqnarray}
Here, $\Omega_{_S}, \Omega_{_B}, \Omega_{_G}$ are the ordinary
density parameters for each matter component (supergravity gas,
brane gas, and gauge field, respectively), and $\Omega_\sigma$ 
the fractional contribution of the shear to the expansion of the 
Universe.
With these definitions the constraint (\ref{eq:Einstein_t}) 
becomes simply,
\begin{equation}
 \Omega_\sigma 
+\Omega_{_S}
+\Omega_{_B}
+\Omega_{_G}
= 1\, .
\label{eq:Constraint}
\end{equation}

\section{Dynamical features at late times}
Let us start the analysis of the late time dynamics by describing
some general properties. 
Without a gas of branes and no gauge degrees of freedom,
flat and Kasner spacetimes are exact vacuum solutions
with $\rho_{_S}=0$.
When the energy density of the gas of supergravity particles
is nonzero, the cosmological dynamics is that of an 
isotropic eleven-dimensional radiation dominated universe
with a growing scale factor $e^\lambda\sim t^{2/11}$.
On the other hand, for $\rho_{_S}=0$ and a gas of branes 
wrapping all the spatial dimensions isotropically an exact
solution exits with $e^\lambda\sim t^{1/4}$ \cite{Easther:2002qk}.
It is easy to show that in general the energy density
of the supergravity particles cannot fully dominate at late
times if a brane gas with a nonisotropic wrapping matrix is
present.
As a consequence, the final fate of the Universe in this case
is always anisotropic and a hierarchy of dimensions can always 
be created. 
Finally, if $\rho_{_G}\neq 0$ and all other energy 
components vanish, one can find an asymptotic power-law solution,
\begin{equation}
e^{\lambda_i}
   \sim \left\{
               \begin{array}{ll}
                 t^{-2/7} & \textrm{\,\,\, for } i=1, 2, 3 \\
                 t^{1/7} & \textrm{\,\,\, for } i=4, \cdots, 10
               \end{array}
         \right.
\end{equation}
This clearly means that the dynamical effect of the form field 
is to drive a cosmological evolution of the Universe with $3$
contracting and $7$ expanding spatial dimensions. 
Then, if the asymptotic evolution of these type of cosmologies 
is fully dominated by the gauge sector, even in the presence of 
wrapping branes, the correct number of large spatial dimensions 
cannot be explained and the present decompactification mechanism 
would be completely invalidated. 
Fortunately, as we will see shortly, this is not the case.
Another important point to stress from this scaling analysis
is that a higher-dimensional Universe dominated by a $4$-form field 
has subvolumes that shrink.
Consequently, unless quantum effects prevent the contracting
dimensions from reaching a zero size, a physical singularity 
appears in the future.
Furthermore, one can find exact solutions in which this 
singularity occurs at a finite proper time \cite{Demaret:1985js}.
Note that, this picture can qualitatively change if the 
underlying spacetime has a submanifold with spatial curvature.

\subsection{Cosmological evolution without fluxes}
First consider the dynamics without fluxes.
It is important to remark that all the comments we are going to 
make in this section are also applicable to those situations in 
which the fluxes are negligible at late times but not necessarily 
at intermediate stages.

We take initially a Universe filled with a gas of supergravity 
particles and a gas of M2-branes described by a wrapping 
matrix with $m_1$ unwrapped, $m_2$ partially wrapped, and 
$m_3=10-m_1-m_2$ fully wrapped spatial dimensions.
Generically, we will refer to such brane configurations as a
brane gas with $m_1$-$m_2$-$m_3$ wrapping.
In general, the solutions of the Einstein equations 
(\ref{eq:Einstein_s}) present an universal power-law behaviour 
at late times.
That is, the sizes of all the spatial dimensions grow with a power
law which depends solely on the wrapping matrix and not on the  
initial conditions \cite{Easther:2002qk}. 
These attractor solutions have a generic analytical form given by,
\begin{equation}
e^{\lambda_i}
   \sim \left\{
               \begin{array}{ll}
                 t^{\alpha} & \textrm{\,\,\, for } i=1, \cdots, m_1 
                              \textrm{\,\, (unwrapped)}\\
                 t^{\beta} &  \textrm{\,\,\, for } i=m_1+1, \cdots, m_1+m_2 
                              \textrm{\,\, (partially wrapped)}\\
                 t^{\gamma} & \textrm{\,\,\, for } i=m_1+m_2+1, \cdots, 
                                                     m_1+m_2+m_3
                              \textrm{\,\, (fully wrapped)}
               \end{array}
         \right.                                           \label{eq:attractor}
\end{equation}
As discussed in \cite{Easther:2002qk}, a physically relevant 
hierarchy of dimensions is produced only if $m_1=3$.
For instance, a brane gas with a wrapping matrix of type 3-3-4 has,
\begin{equation}
\alpha = \frac{7}{13}\, ,
\,\,\,\,\,\,\,
\beta = \gamma = \frac{1}{13}\, ,                              \label{eq:nf334}
\end{equation}
and another with wrapping matrix of type 3-4-3, 
\begin{equation}
\alpha = \frac{8}{15}\, ,
\,\,\,\,\,\,\,
\beta = \frac{2}{15}\, ,
\,\,\,\,\,\,\,
\gamma = 0\, .                                                 \label{eq:nf343}
\end{equation}
These two examples illustrate that a gas of branes with the
appropriate wrapping can support naturally a faster growth 
of three spatial dimensions because $\alpha$ is always greater
than $\beta$ and $\gamma$. 
This is because the brane gas exerts no pressure on unwrapped 
directions and, consequently, cannot suppress their expansion.
For partially and fully wrapped dimensions the pressures
are negative and cause the opposite effect.
One also observes, that there exist configurations in which 
$\beta > \gamma$ and a sub-hierarchy in the sizes of small 
directions can also be formed.
This feature occurs in configurations with wrapping parameters
satisfying $m_2>m_3$.
In fact, one can further show that the powers only depend on the 
total number of spatial dimensions (in our case $10$) and the 
wrapping matrix parameters $m_1, m_2,$ and $m_3$
(the explicit expressions can be found in \cite{Easther:2002qk}).

An account of the full dynamics can be obtained by solving 
Einstein equations numerically (see Fig~\ref{fig:nofluxes}).
Since we do not expect supergravity to hold at very early times,
the strategy we have followed is to evolve the system from the 
Planck time onwards given initial random values for $\lambda_i$ 
and $\dot\lambda_i$.
For the nonzero components of the wrapping matrix $N_{ij}$ we
take random integers chosen from the set $\{1, 2, 3\}$.
We have checked for consistency that the final expansion rates 
for each spatial dimension are independent of these choices.
The constraint (\ref{eq:Einstein_t}), or (\ref{eq:Constraint}),
is only used to compute the initial value of the energy density 
of the supergravity gas, $\rho^o_{_S}$, and to check the numerical 
accuracy of the output. 
In our particular numerical implementation, the universal powers 
found in all the examples agree with those obtained analytically 
with less than $1\%$ error.
In general, the dynamics presents three stages.
A short initial phase in which any initial anisotropic expansion
is rapidly diluted, a much longer intermediate phase of isotropic 
expansion, and a final phase in which the unwrapped dimensions grow 
with an expansion rate larger than that of the rest of spatial 
dimensions.
As one can observe from the plots depicting the contributions
to the constraint (\ref{eq:Constraint}), the shear, which represents 
the degree of anisotropy of the Universe, decreases in the first phase, 
is negligible in the second, and grows again until it scales with
the mean expansion in the third.
The energy density of the supergravity particles dominates the
expansion of the Universe in the first two stages and the energy 
density of the brane gas in the final stage. 
The time at which both contributions are of the same magnitude 
marks the transition between the last two phases.
For the simulations represented in Fig.~\ref{fig:nofluxes} this 
time is approximately $t\sim 10^6$ but the actual number depends 
on the initial conditions and the wrapping matrix chosen.

\FIGURE[t]{
\epsfig{file=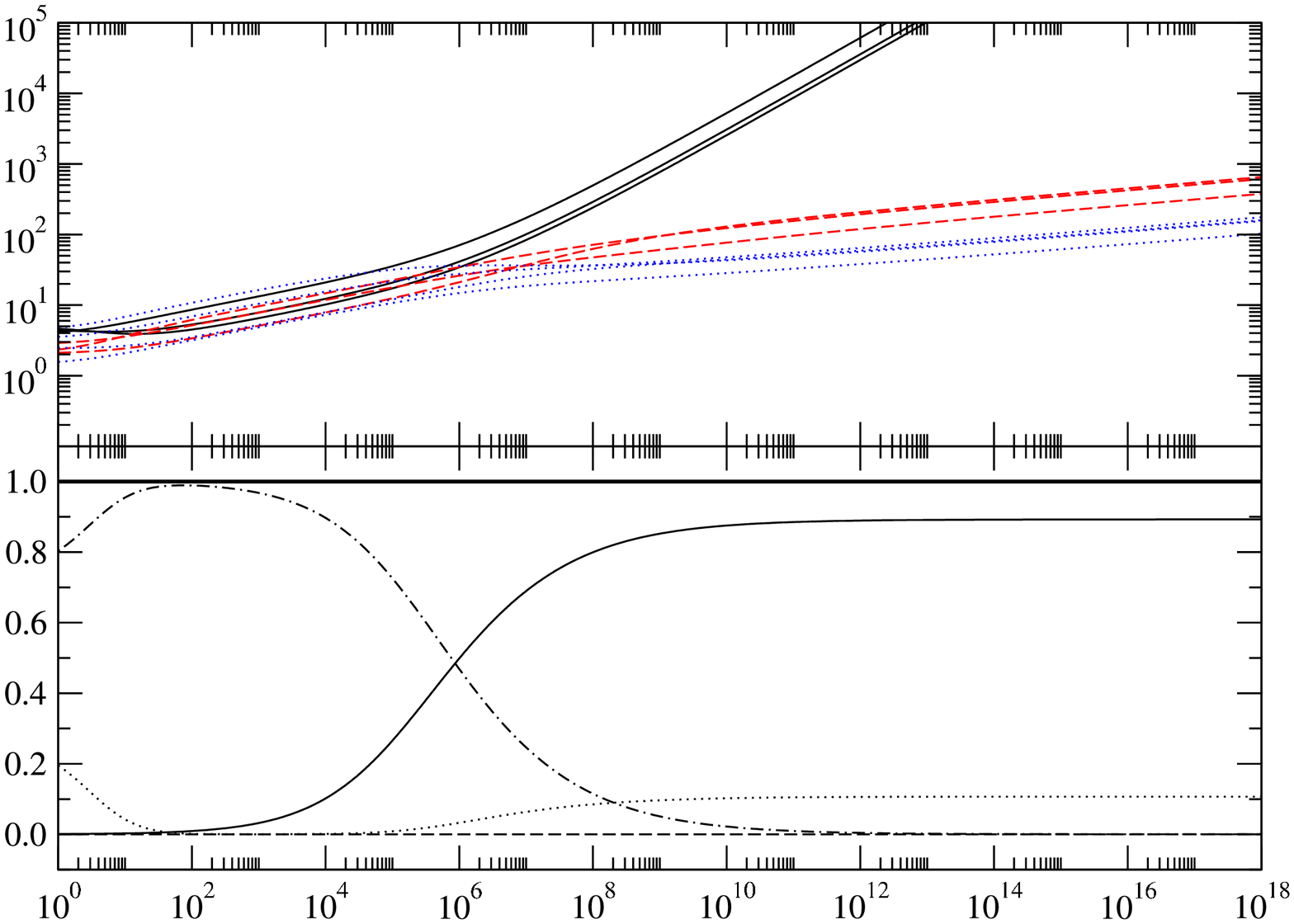,height=4in,width=0.49\columnwidth}
\epsfig{file=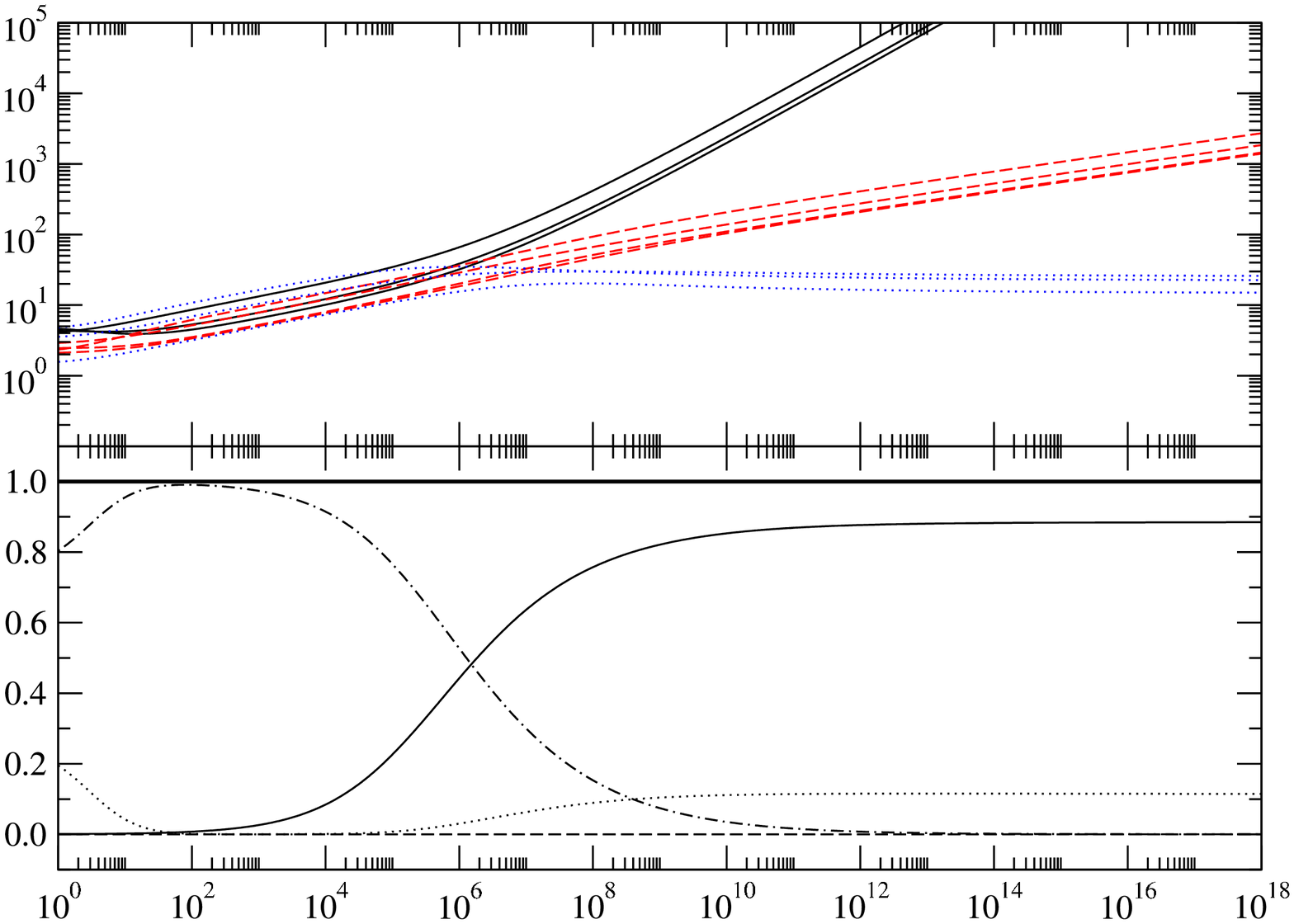,height=4in,width=0.49\columnwidth}
\caption{
Cosmological evolution without fluxes. 
Graphs on the left are for a brane gas with a wrapping matrix of 
the type 3-3-4 and those on the right of the type 3-4-3. 
Plots on top represent the time evolution of the size (scale factor
normalised by $2\pi$) of all the spatial dimensions: 
solid curves are for unwrapped dimensions, dashed curves for partially 
wrapped dimensions, and dotted curves for fully wrapped dimensions. 
Plots on the bottom depicts all the contributions to the expansion
of the Universe as a function of cosmic time. The solid line represents 
$\Omega_{_B}$, the dashed line $\Omega_{_G}$, the dotted line 
$\Omega_\sigma$, and the dotted-dashed line $\Omega_{_S}$. 
The thick solid line is the sum of all the contributions and serves to 
check the accuracy of the numerical computation (\ref{eq:Constraint}).
\label{fig:nofluxes}}
}

In conclusion, the cosmology of brane gases in a (low-energy)
M theory context opens the possibility of explaining the number 
of spatial dimensions observed today for those configurations
characterised by a wrapping parameter $m_1=3$.
In the following we see that this conclusion is still true
even if fluxes have a significant contribution at late times.
Furthermore, we present new cosmological solutions with a 
larger number of initially unwrapped directions which support
asymptotically a large four dimensional spacetime. 

\subsection{Cosmological evolution with fluxes}
Now we are interested to investigate the influence of
fluxes on the late-time cosmological dynamics of a Universe
filled with a gas of M2-branes.
The first difficulty one faces when fluxes are present is
that the wrapping matrix describing the branes gas and the
gauge field induce different splittings of the spacetime.
As we have seen, the type of solutions for the 4-form field 
strength we are considering naturally separates the spacetime 
into R$\times$T$^3$$\times$T$^7$ and the gas of M2-branes into 
R$\times$T$^{m_1}$$\times$T$^{m_2}$$\times$T$^{m_3}$.
In spite of that, the universal power-law scaling behaviour at 
late times we have described in the previous section, although 
slightly modified, is not lost.  

Given a brane gas configuration with wrapping matrix of type
$m_1$-$m_2$-$m_3$ the late cosmological evolution can be cast into
the form,
\begin{equation}
e^{\lambda_i}
   \sim \left\{
               \begin{array}{ll}
                 t^{\alpha_-} & \textrm{\,\,\, for } i=1, \cdots, 3 
                                \textrm{\,\, (unwrapped)}\\
                 t^{\alpha_+} & \textrm{\,\,\, for } i=4, \cdots, m_1 
                                \textrm{\,\, (unwrapped)}\\
                 t^{\beta} &    \textrm{\,\,\, for } i=m_1+1, \cdots, m_1+m_2 
                                \textrm{\,\, (partially wrapped)}\\
                 t^{\gamma} &   \textrm{\,\,\, for } i=m_1+m_2+1, \cdots, 
                                                     m_1+m_2+m_3
                              \textrm{\,\, (fully wrapped)}
               \end{array}
         \right.                                    \label{eq:attractor_fluxes}
\end{equation}
where we have chosen the first three directions to be those picked 
by the Freund-Rubin ansatz for the gauge field.  
As seen in Fig.~\ref{fig:fluxes_3}, for configurations with a
wrapping matrix 3-$m_2$-$m_3$, the energy density of the
gauge field and the brane gas both have a significant 
contribution at late times.
In this case (note that there are no directions with $\alpha_+$) 
the scaling behaviour is given analytically by,
\begin{equation}
\alpha_- = \frac{3}{7}\, ,
\,\,\,\,\,\,\,
\beta = \gamma = \frac{1}{7}\, ,
\end{equation}
when $m_2 < m_3$, and,
\begin{equation}
\alpha_- = \frac{5m_2m_3-14}{5m_2m_3+49}\, ,
\,\,\,\,\,\,\,
\beta = \frac{5m_3(m_2-3)+7}{5m_2m_3+49}\, ,
\,\,\,\,\,\,\,
\gamma = \frac{5m_2(m_3-3)+7}{5m_2m_3+49}\, ,
\end{equation} 
when $m_2 > m_3$.
As a particular example of the last case, one has
\begin{equation}
\alpha_- = \frac{46}{109}\, ,
\,\,\,\,\,\,\,
\beta = \frac{22}{109}\, ,
\,\,\,\,\,\,\,
\gamma = \frac{7}{109}\, ,
\end{equation} 
for $m_2=4$ and $m_3=3$.
Comparing these results with (\ref{eq:nf334}) and (\ref{eq:nf343}) 
one readily observes that, although in both types of configurations 
the expansion rates of the unwrapped directions are decreased and 
those of the partially and fully wrapped  are increased, {\sl the 
influence of fluxes at late times does not destroy the hierarchy 
between large and small dimensions} ($\alpha_-$ is still greater
than $\beta$ and $\gamma$ in all the cases).
Note as well that even though the energy density of the gauge
field fully dominates at early times, the energy density of
the gas of branes very quickly takes over and the collapse
of three dimensions is avoided.  
The appearance of a physical singularity in this early 
phase of the evolution is then prevented at the classical level.

\FIGURE[t]{
\epsfig{file=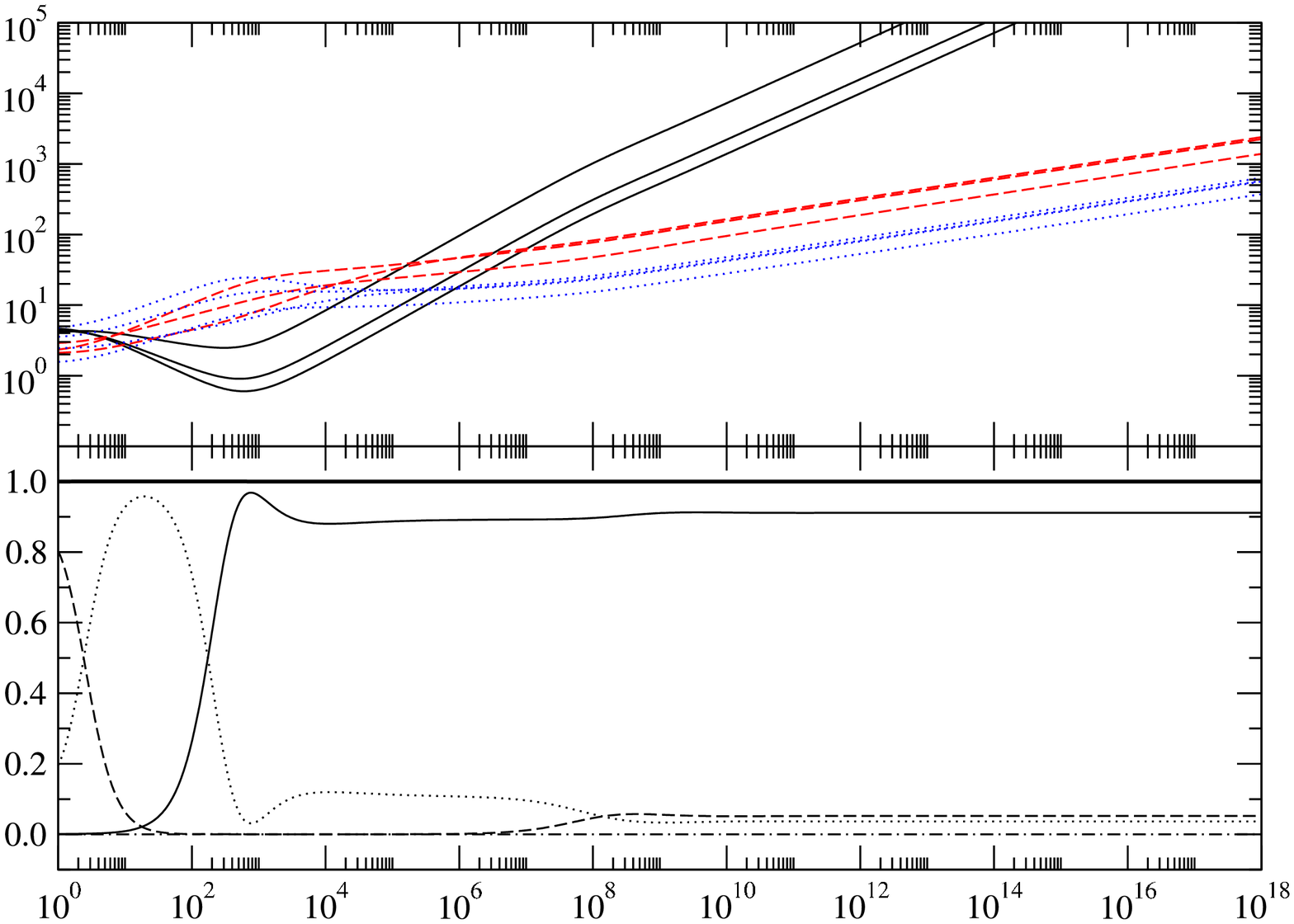,height=4in,width=0.49\columnwidth}
\epsfig{file=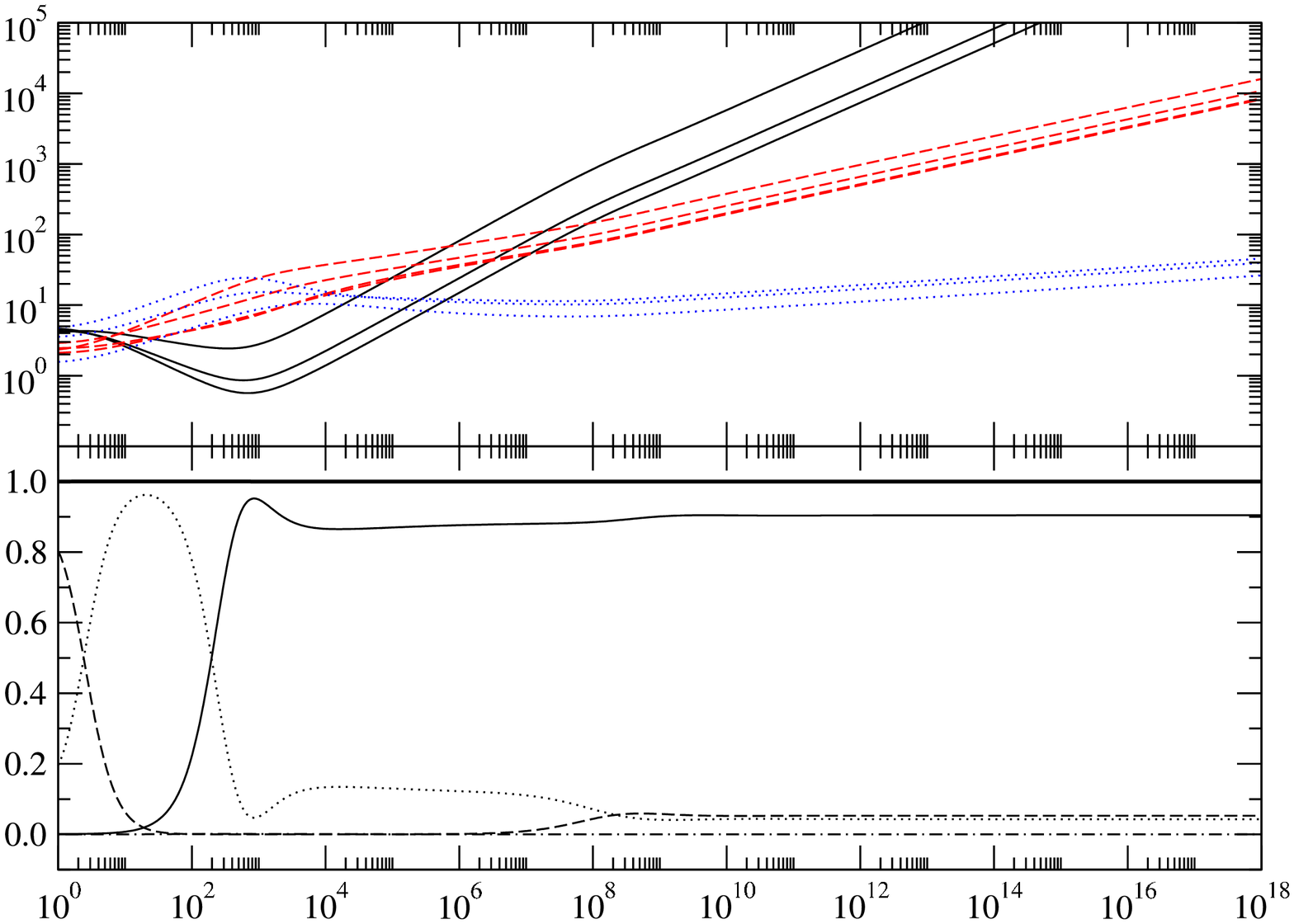,height=4in,width=0.49\columnwidth}
\caption{
Cosmological dynamics of a brane gas of wrapping type 
3-3-4 (left) and type 3-4-3 (right) with fluxes. 
The identification of plots and lines is the same as in 
Fig.~\ref{fig:nofluxes}.
\label{fig:fluxes_3}}
}

A brane configuration of particular interest is that of a brane gas 
with 6 unwrapped dimensions; that is, configurations with wrapping
matrix of type 6-$m_2$-$m_3$.
Without fluxes, there are only four possible configurations leading 
to an anisotropic evolution. 
For all the configurations with $m_2 \leq m_3$ one has the analytic 
solution (\ref{eq:attractor}) with,
\begin{equation}
\alpha = \frac{4}{11}\, ,
\,\,\,\,\,\,\,
\beta = \gamma = -\frac{1}{11}\, .                             \label{eq:nf6mm}
\end{equation}
On the other hand, for the configuration with wrapping 6-3-1,
which is the only one with $m_2 > m_3$ supporting a 
nonisotropic expansion, one obtains the scaling,
\begin{equation}
\alpha = \frac{23}{77}\, ,
\,\,\,\,\,\,\,
\beta = \frac{8}{77}\, ,
\,\,\,\,\,\,\,
\gamma = -\frac{22}{77}\, .                                    \label{eq:nf631}
\end{equation}
Obviously, if the dynamical effects of the gauge field are
not taking into account these configurations, although 
providing a hierarchy among different dimensions,
do not predict the right number of large spatial directions.
Finally, it is important to note that, contrary to what happens 
when $m_1=3$, the energy density of the supergravity particles 
is always not negligible at late times.

Let us now consider the dynamical effects of the gauge degrees 
of freedom in configurations with $m_1=6$.
Two numerical examples are compared with the no flux case 
in Figs.~\ref{fig:613_EL}~and~\ref{fig:631_EL}. 
An interesting point is that all the energy components 
($\Omega_{_B}, \Omega_{_S}, \Omega_{_G}$) have a significant
contribution to the cosmological expansion at late times.
As one can also observe, these solutions represent a new family
of configurations that permits the growth of three spatial 
dimensions.
Although for the configuration with wrapping matrix 6-3-1
the hierarchy of the larger dimensions is not quite obvious 
from the plot obtained numerically (Fig.~\ref{fig:631_EL}), 
it can be seen analytically that the asymptotic behaviour of
the solution has the form (\ref{eq:attractor_fluxes}) with,
\begin{eqnarray}
&&\alpha_- = \frac{12}{44}\, ,
\,\,\,\,\,\,\,
\beta = \frac{5}{44}\, ,
\nonumber\\
&&\alpha_+ = \frac{14}{44}\, ,
\,\,\,\,\,\,\,
\gamma = -\frac{13}{44}\, .
\end{eqnarray} 
The values obtained in our numerical computation agree with the 
above analytical result up to four decimal places.
Comparing with the case without fluxes, $\beta$ is larger and
$\gamma$ smaller.
That means that the three partially wrapped dimensions grow faster
and the fully wrapped dimension contracts also faster.
On the other hand, the six unwrapped dimensions are separated
into two groups, of three dimensions each, with independent 
expansion rates defined by $\alpha_-$ and $\alpha_+$, respectively. 
Since $\alpha_- < \alpha_+$ the appropriate hierarchy
for explaining the number of spatial dimensions is produced. 
On the other hand, for $m_2 \leq m_3$ one finds analogously the 
asymptotic power-laws (see Fig.~\ref{fig:613_EL}),
\begin{equation}
\alpha_- = \frac{3}{11}\, ,
\,\,\,\,\,\,\,
\alpha_+ = \frac{5}{11}\, ,
\,\,\,\,\,\,\,
\beta = \gamma = -\frac{1}{11}\, .
\end{equation}
Again $\alpha_+ > \alpha_- > \beta, \gamma$ and a hierarchy
with three dimensions growing larger is obtained.
Consequently, {\sl brane configurations with $m_1=6$ unwrapped 
dimensions can naturally explain why three large spatial 
dimensions are decompactified at late times if the gauge sector 
of eleven-dimensional supergravity is turned on.}
This new family of solutions enlarge the possible wrapping 
configurations at the end of brane-antibrane annihilation in 
the Hagedorn phase. 
It is important to emphasis that the interplay between both the 
brane gas and the flux dynamics plays a fundamental role in 
getting the correct number of large dimensions.

\FIGURE[t]{
\epsfig{file=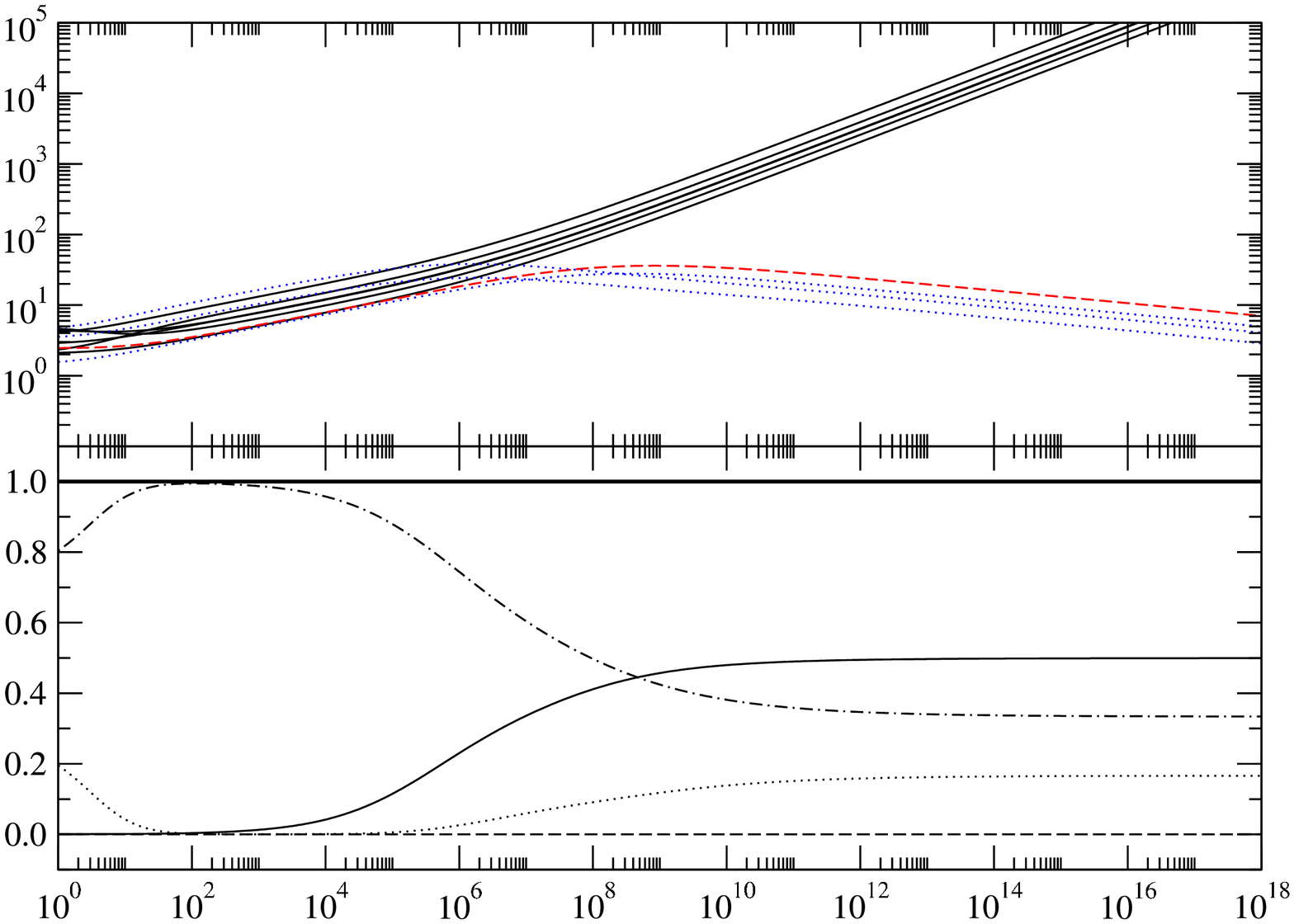,height=4in,width=0.49\columnwidth}
\epsfig{file=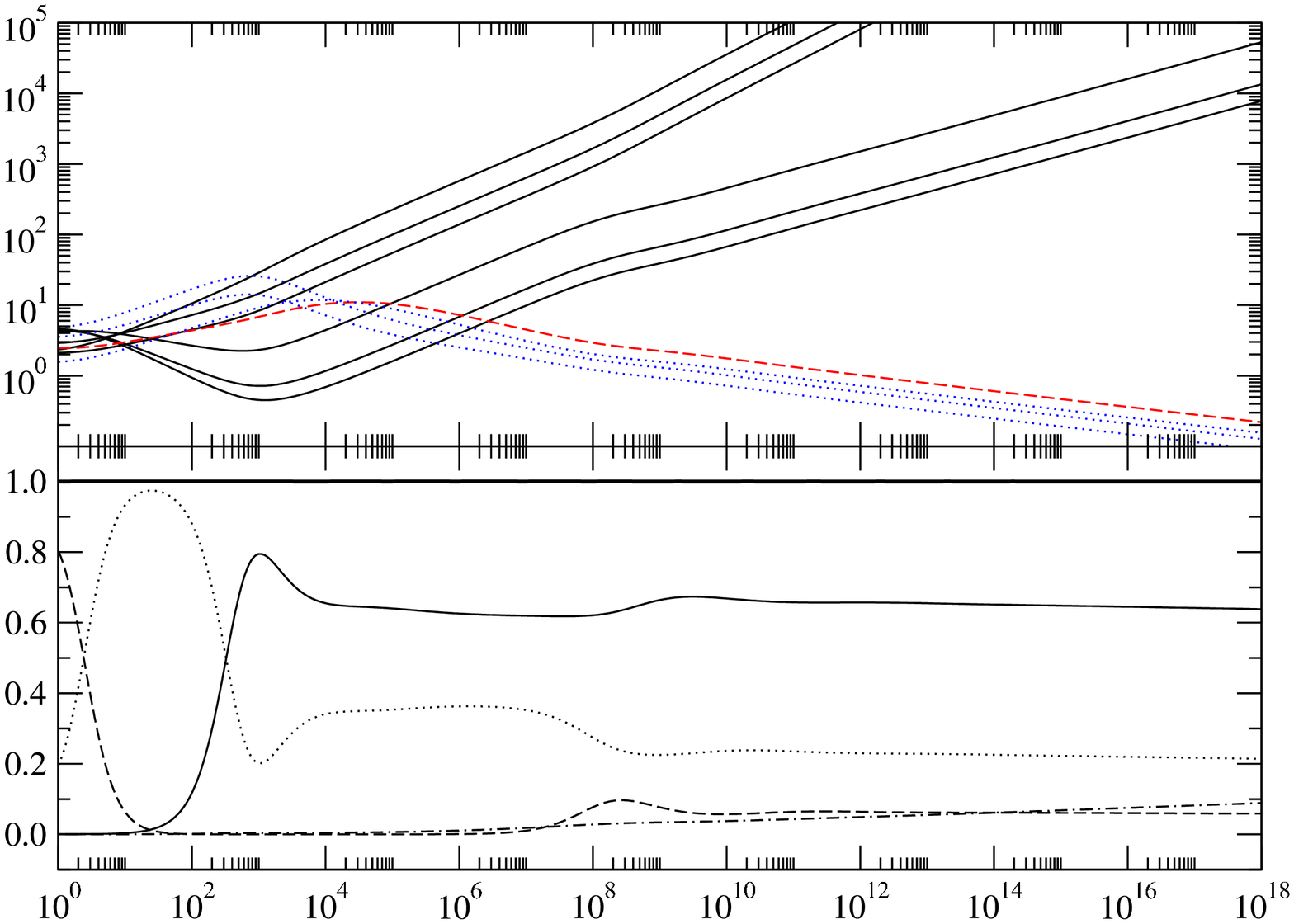,height=4in,width=0.49\columnwidth}
\caption{
Cosmological dynamics of a brane gas of wrapping type 
6-1-3 without (left) and with (right) fluxes. 
The identification of plots and lines is the same as in 
Fig.~\ref{fig:nofluxes}.
\label{fig:613_EL}}
}

\FIGURE[t]{
\epsfig{file=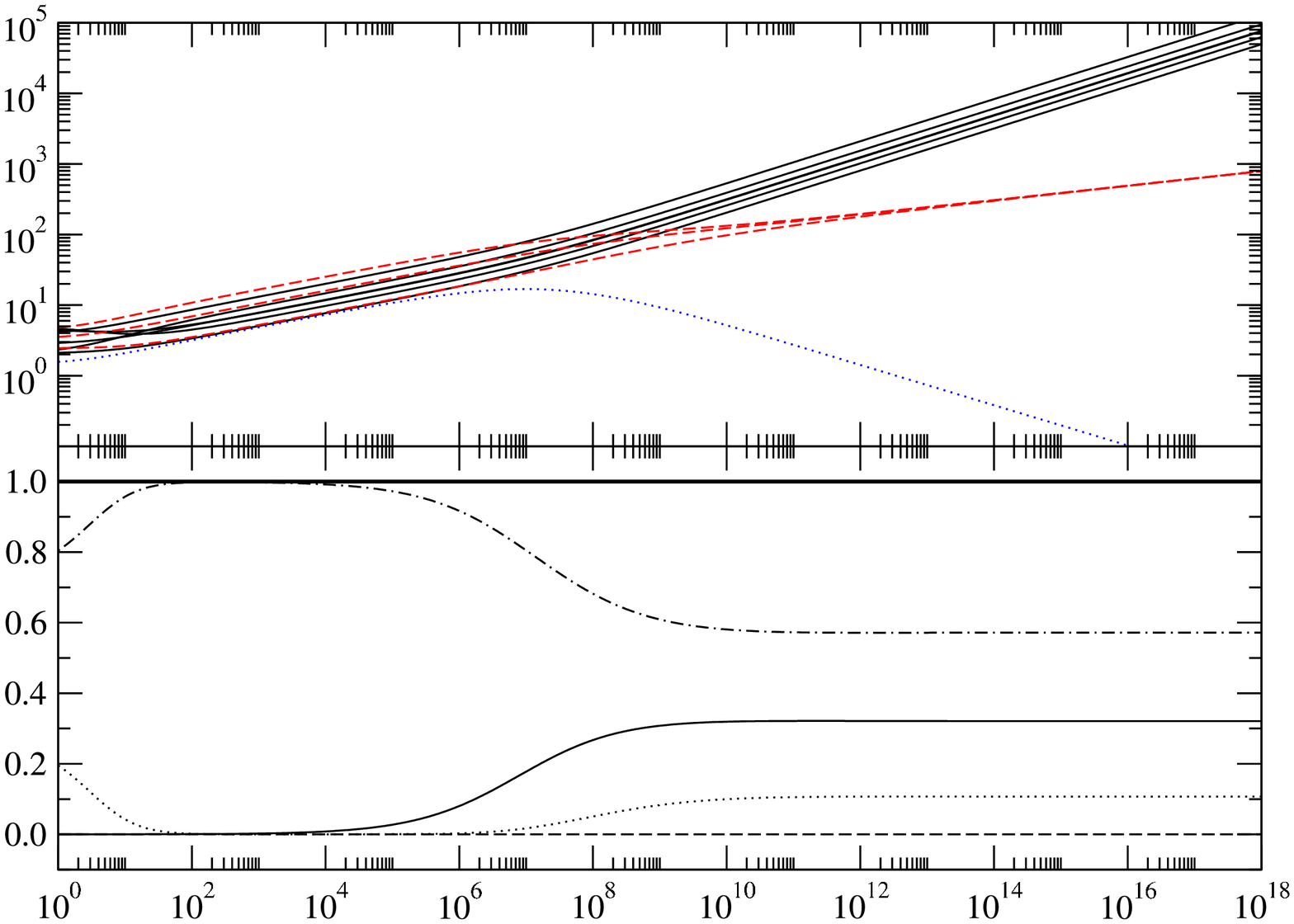,height=4in,width=0.49\columnwidth}
\epsfig{file=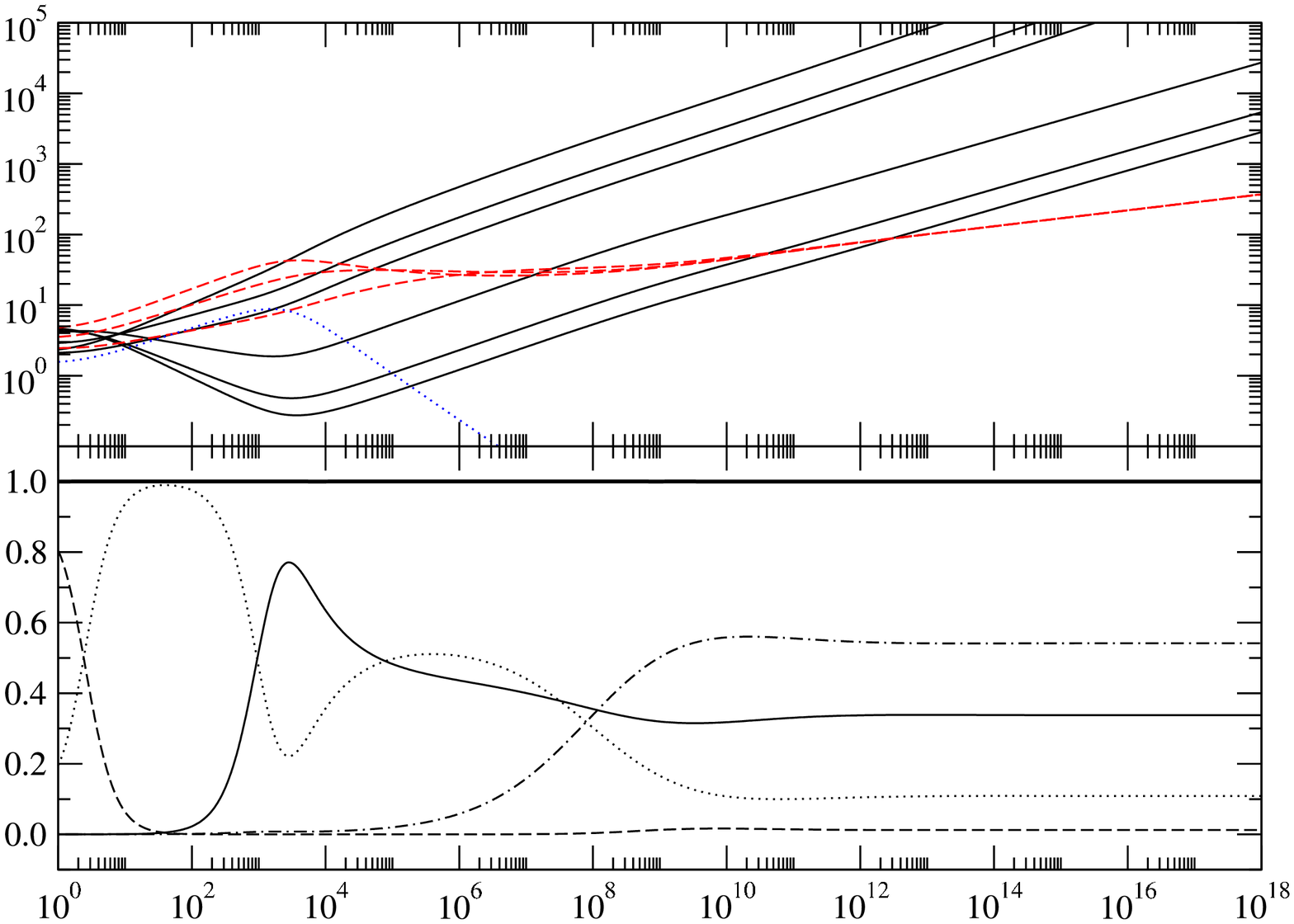,height=4in,width=0.49\columnwidth}
\caption{
Cosmological dynamics of a brane gas of wrapping type 
6-3-1 without (left) and with (right) fluxes. 
The identification of plots and lines is the same as in 
Fig.~\ref{fig:nofluxes}.
\label{fig:631_EL}}
}

\section{Conclusions}
We have illustrated how the cosmological evolution at late time 
of a gas of M2-branes within a low-energy limit of M theory is
modified when the gauge fields of the bosonic sector are taking
into account.

We have seen that fluxes respect the hierarchies among
different spatial dimensions introduced by an anisotropically 
wrapped brane gases at late times.
We have also found new solutions that can explain the actual
number of spatial dimensions of the Universe which are 
characterised by a brane gas configuration with a large 
number of unwrapped dimensions.
These solutions appear as far as the gauge field strength
is sufficiently strong to have a significant contribution
to the total cosmological expansion at late times. 
On thermodynamical grounds one should expect that 
these brane configurations can be originated at the
end of the Hagedorn phase from smaller initial spatial
volumes of the Universe than configurations with lower 
unwrapping numbers.
Including gauge fields into the dynamics increases the possible 
initial conditions for the Universe and then the probability of 
obtaining decompactification from anisotropically wrapped 
spacetimes.  
This makes less severe the fine-tuning problem posed 
by this mechanism when fluxes are not considered.

A difficult issue is to assess whether the constraints
on the initial size of the Universe imposed by the holographic
principle are significatively modified when gauge degrees of
freedom are included.
For the type of solutions we have studied the assumption 
of isotropy as considered in \cite{Easther:2003dd} is certainly 
not acceptable and a fully anisotropic analysis of entropy 
bounds is absolutely compulsory.
In principle, the presence of a new physical parameter 
representing the strength of the gauge field opens the
possibility of getting less restrictive conditions on the
physical volume of the Universe in the Hagedorn phase.

Unfortunately, as in the case without fluxes, the internal 
dimensions do not stabilise for the physically interesting 
cases. 
In general, it seems rather difficult to get stabilisation
within this mechanism without introducing additional physics. 

Finally, it would be certainly interesting to investigate if 
there exist solutions with an even larger number of unwrapped 
dimensions that could as well explain the number of spacetime 
dimensions.
A full classification of solutions and configurations will 
determine how generic is this anisotropic mechanism of 
decompactification.
This analysis will further reveal the connection with the 
mechanism suggested in \cite{Campos:2003ip} within the context 
of ten-dimensional type IIA supergravity. 
This lower dimensional theory arises as a result of the 
compactification of eleven-dimensional supergravity and,
thus, in principle, one should expect a close relationship
between the cosmological solutions in both frameworks.

\acknowledgments

The author thanks M. Seco for valuable discussions on the numerical
approach and also acknowledges the support of the Alexander von 
Humboldt Stiftung/Foundation and the Universit\"at Heidelberg.

\bibliography{bgc}

\end{document}